\documentclass[reprint,aps,prl,amsmath,amssymb,superscriptaddress,floatfix]{revtex4-1}

\usepackage{booktabs}
\usepackage{graphicx}
\usepackage{dcolumn}
\usepackage[colorlinks=true,linkcolor=blue,anchorcolor=blue, citecolor=blue,urlcolor=blue]{hyperref}
\usepackage[mathlines]{lineno}

\begin{document}
\title{Demonstration of High-Efficiency Microwave Heating Producing Record Highly Charged Xenon Ion Beams with Superconducting ECR Ion Sources}

\author{X. Wang}
\author{J. B. Li}
\email{jiboli@impcas.ac.cn}
\affiliation{Institute of Modern Physics, Chinese Academy of Sciences, Lanzhou 730000, China}
\affiliation{School of Nuclear Science and Technology, University of Chinese Academy of Sciences, Beijing 100049, China}

\author{V. Mironov}
\affiliation{Joint Institute for Nuclear Research, Flerov Laboratory of Nuclear Reactions, Dubna, Moscow Region 141980, Russia} 

\author{J. W. Guo}
\affiliation{Institute of Modern Physics, Chinese Academy of Sciences, Lanzhou 730000, China}

\author{X. Z. Zhang}
\affiliation{Institute of Modern Physics, Chinese Academy of Sciences, Lanzhou 730000, China}
\affiliation{School of Nuclear Science and Technology, University of Chinese Academy of Sciences, Beijing 100049, China}

\author{O. Tarvainen}
\affiliation{UKRI Science and Technology Facilities Council, Rutherford Appleton Laboratory, Harwell Campus, Oxfordshire, OX110QX, U.K.}

\author{Y. C. Feng}
\author{L. X. Li}
\author{J. D. Ma}
\author{Z. H. Zhang}
\affiliation{Institute of Modern Physics, Chinese Academy of Sciences, Lanzhou 730000, China}

\author{W. Lu}
\affiliation{Institute of Modern Physics, Chinese Academy of Sciences, Lanzhou 730000, China}
\affiliation{School of Nuclear Science and Technology, University of Chinese Academy of Sciences, Beijing 100049, China}

\author{S. Bogomolov}
\affiliation{Joint Institute for Nuclear Research, Flerov Laboratory of Nuclear Reactions, Dubna, Moscow Region 141980, Russia}

\author{L. Sun}
\email{sunlt@impcas.ac.cn}
\author{H. W. Zhao}
\affiliation{Institute of Modern Physics, Chinese Academy of Sciences, Lanzhou 730000, China}
\affiliation{School of Nuclear Science and Technology, University of Chinese Academy of Sciences, Beijing 100049, China}

\date{\today}

\begin{abstract}
Intense highly charged ion beam production is essential for high-power heavy ion accelerators. A novel movable Vlasov launcher for superconducting high charge state Electron Cyclotron Resonance (ECR) ion source has been devised that can affect the microwave power effectiveness by a factor of about 4 in terms of highly charged ion beam production. This approach based on a dedicated microwave launching system instead of the traditional coupling scheme has led to new insight on microwave-plasma interaction. With this new understanding, the world record highly charged xenon ion beam currents have been enhanced by up to a factor of 2, which could directly and significantly enhance the performance of heavy ion accelerators and provide many new research opportunities in nuclear physics, atomic physics and other disciplines. 
\end{abstract}

\maketitle

\clearpage

\section{Introduction.}
As the most powerful machine to produce intense highly charged ion beams, Electron Cyclotron Resonance (ECR) ion source \cite{Geller_book} has played an indispensable role in accelerator based nuclear, atomic and material physics as well as various applications, such as exotic nuclei investigation, new elements synthesis, new material discoveries and heavy ion cancer treatment. The increasing beam intensity need of high-power heavy ion accelerators requests a significant performance enhancement of ECR ion source \cite{Zhou2022,doi:10.1142/S0218301319300030,franbergdelahaye:hiat2022-we2i1,Okuno_2020}, whereas over the last 20 years, there have been no obvious progresses towards new techniques and interpretation on microwave-plasma interaction of a high performance ECR ion source. Our recent work as discussed in this paper may be a breakthrough to the advancement of ECR ion source and boost the performance of the state-of-the-art machines in operation, which provides new possibilities to the physics goals as well as heavy ion accelerator facility performance, such as the $400~\text{kW}$ goal with FRIB accelerator \cite{PhysRevLett.126.114801}, and the $10\sim20~\text{p}\mu\text{A}$ $^{54}\text{Cr}^{14+}$ beam for 119, 120 elements synthesis \cite{PhysRevC.102.064602}.

The ECR ion source is a device with a magnetically confined plasma sustained by microwave radiation. Electrons in the plasma are heated through electron cyclotron resonance heating (ECRH) to high energies by the coupled microwave power, and highly charged ions are produced by electron impact ionization. Thus, the performance of the ECR ion source is directly influenced by the efficiency of coupling the microwave energy to the plasma electrons. In general, for microwave frequencies $f\leq 18~\text{GHz}$, most ECR ion sources adopt a rectangular waveguide excited with a single $\text{TE}_{10}$ mode~\cite{10.1063/1.3265366}. This approach works well for power levels up to $2~\text{kW}$~\cite{Hitz2006RecentPI}, corresponding to the maximum output power of an $18~\text{GHz}$ klystron amplifier. For microwave frequencies $f>20~\text{GHz}$, a gyrotron is used to generate the microwave power up to $10~\text{kW}$. Following the first successful injection of $28~\text{GHz}$ microwaves into the SERSE ion source~\cite{Gammino2002OperationsOT}, all subsequent $3^{rd}$ generation ECR ion sources~\cite{10.1063/1.1893402,10.1063/1.4825164,Riken}, which represent the state-of-the-art of high charge state ECR ion source technologies and performances, use an oversized circular waveguide ($\Phi=32~\text{mm}$) to launch $\text{TE}_{01}$ mode microwave radiation into the plasma. However, the coupling efficiency at frequencies above $20~\text{GHz}$ is questionable, since it has been verified experimentally~\cite{Hitz2006RecentPI} that at a fixed microwave power level of gyrotron frequencies, the ion source performance fails to follow the theoretical prediction by the `frequency scaling laws' (beam intensity $I\propto f^2$)~\cite{Geller_book}. The less-than-expected microwave heating efficiency severely restricts performance of modern superconducting ECR ion sources and thus becomes an urgent issue for advanced heavy ion accelerators. Moreover, without remarkable progress, this conundrum would become a more severe problem for the next generation ECR ion source operating at an even higher frequency ($f>40~\text{GHz} $)~\cite{10.1063/1.5017479}.

\begin{figure*}[!ht]
    \centering
    \includegraphics[width=\linewidth]{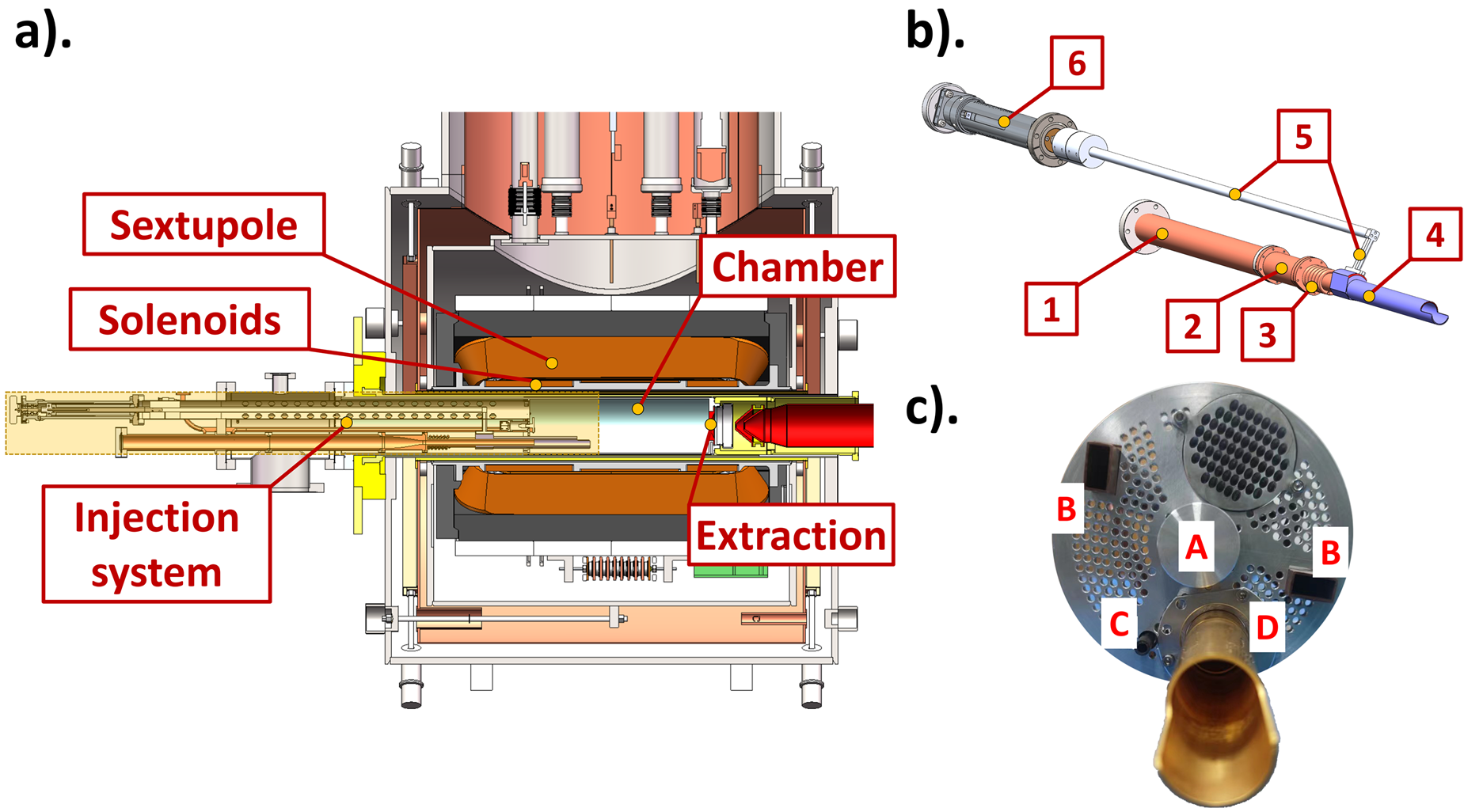}
    \caption{ (a) The schematic drawing of the SECRAL-II ion source. (b) The schematic diagram of the movable Vlasov launcher, consisting of a standard circular waveguide (1), a $32~\text{mm}$ to $20~\text{mm}$ transition (2), a water-cooled circular waveguide (3), a Vlasov launcher (4), a mechanical linkage (5) and a ball screw cylinder (6). (c) Plasma chamber injection system of SECRAL-II ion source. The biased disc (A) is at the center, surrounded on both sides by two WR-62 type rectangular waveguides (B). The gas inlet (C) is at the bottom. The movable Vlasov launcher is located in place (D).}
    \label{fig:Structure-Outline}
\end{figure*}

Continuous efforts over the past decade have aimed at enhancing the coupling efficiency of gyrotron frequency heating. Injection of the microwaves in the HE$_{11}$ mode was firstly proposed~\cite{Hitz2006RecentPI} for better ECRH because of the Gaussian spatial profile of that particular EM-wave mode, but so far two experimental explorations have failed to demonstrate the expected improvements~\cite{10.1063/1.4832064,Sun2016AdvancementOH}. Later on researchers at IMP (Institute of Modern Physics, Chinese Academy of Sciences) found that a circular waveguide with smaller diameter ($\Phi=16\sim20~\text{mm}$) could improve the ion source performance at the same microwave power level. The qualitative interpretation suggests that this enhancement results from optimizing the power distribution on the ECR surface~\cite{10.1063/1.5131101}. This observation gives a hint that an asymmetric Vlasov launcher~\cite{10.1007/BF01037072} would be more suitable to control the microwave power distribution on the ECR surface. Indeed, preliminary experiments demonstrated that the beam current could be increased with a $\text{TE}_{01}$ mode Vlasov launcher~\cite{10.1063/1.5131101}. The position of the Vlasov launcher with respect to the plasma chamber was fixed in these experiments, and it was not possible to systematically investigate or optimise the effect of the power distribution on the ECR surface. To investigate further what the optimal way of injecting the microwaves with frequency above $20~\text{GHz}$ into the plasma is, a novel movable Vlasov launcher has been developed at IMP. Here we report the experimental and numerical researches on optimizing the microwave coupling of a superconducting ECR ion source with the new launcher.

\section{Experimental setup.}
The experimental data discussed hereafter are taken with the SECRAL-II (Superconducting ECR ion source with Advanced design in Lanzhou No.~II) ion source. The  schematic drawing of the ion source is shown in~\autoref{fig:Structure-Outline}(a) and comprehensive description can be found in Ref.~\cite{10.1063/1.5017479}.

The schematic diagram of the movable Vlasov launcher is shown in~\autoref{fig:Structure-Outline}(b). A standard circular waveguide (inner diameter $\Phi=32~\text{mm}$) is connected to an $8~\text{kW}$ GyCOM\textsuperscript{\textregistered} $24~\text{GHz}$ gyrotron with the other end of this waveguide connected to a smaller water-cooled circular waveguide (inner diameter $\Phi=20~\text{mm}$) through a transition piece. Then a Vlasov launcher (refer to supplementary material, Section I) with an inner diameter of $23.6~\text{mm}$ is attached to the water-cooled waveguide ensuring  good thermal and electrical contacts. The online movement of the Vlasov launcher is driven by a ball screw cylinder through a mechanical linkage without compromising the vacuum or turning off the microwave power or extraction voltage. The movable Vlasov launcher is inserted into the source chamber through an RF injection port at the SECRAL-II injection flange. The movable Vlasov launcher and relevant parts at the plasma chamber injection system are shown in~\autoref{fig:Structure-Outline}(c).

In this study, the position of the Vlasov launcher is varied from $0~\text{mm}$, corresponding to an axial distance of $64~\text{mm}$ between the tip of the Vlasov launcher and the injection flange, to $83~\text{mm}$, the range being limited by the mechanical structure. One of the WR-62 type rectangular waveguides is connected to an $18~\text{GHz}$ klystron amplifier as the supplemental heating. Both the WR-62 waveguide and the movable Vlasov launcher are equipped with vacuum windows and high voltage breaks. The source extraction voltage is $20~\text{kV}$. In all measurements, the ion source is operated with xenon and oxygen buffer gas and tuned for high charge state xenon ion production. The ion beam currents are measured directly across a $1~\text{k}\Omega$ resistor connected to the Faraday cup~\cite{PhysRevAccelBeams.25.063402,Sun2015STATUSRO}. In addition, to investigate the influence of the Vlasov launcher position on the plasma electrons, the axial bremsstrahlung spectra are synchronously measured, the detailed description of the bremsstrahlung detection system can be found in Ref.~\cite{WOS:000560067700001}.

\begin{figure}[!ht]
    \centering
    \includegraphics[width=\linewidth]{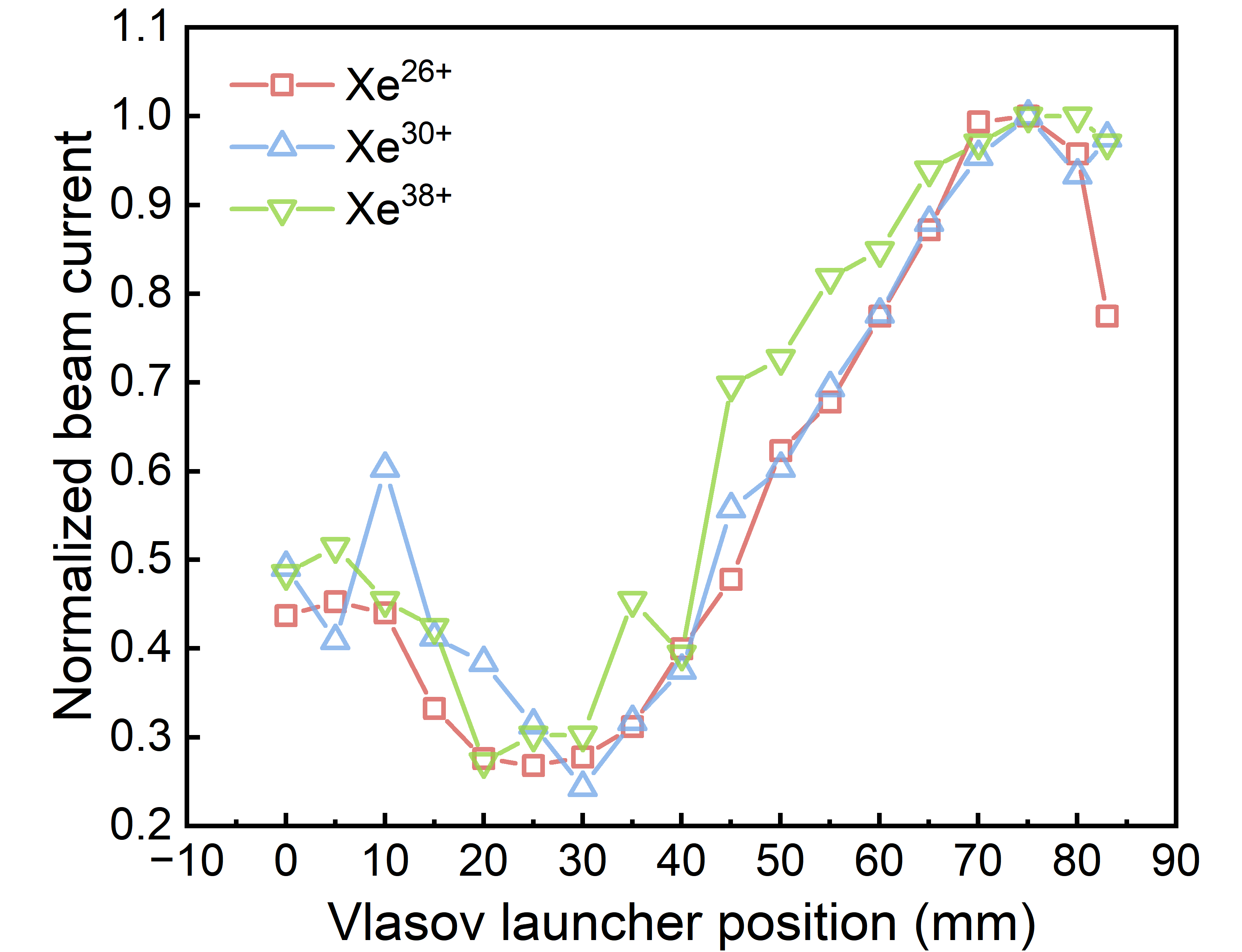}
    \caption{The normalized beam intensities of different charge state xenon beams as a function of the Vlasov launcher position. The microwave power of $24~\text{GHz}$ and $18~\text{GHz}$ are $4.5$ and $0.5~\text{kW}$, respectively. All beam currents in $\mu\text{A}$ units are tabulated in the supplementary material, Table~S1.}
    \label{fig:Experimetal_Results_1}
\end{figure}

\begin{figure}[!ht]
    \centering
    \includegraphics[width=\linewidth]{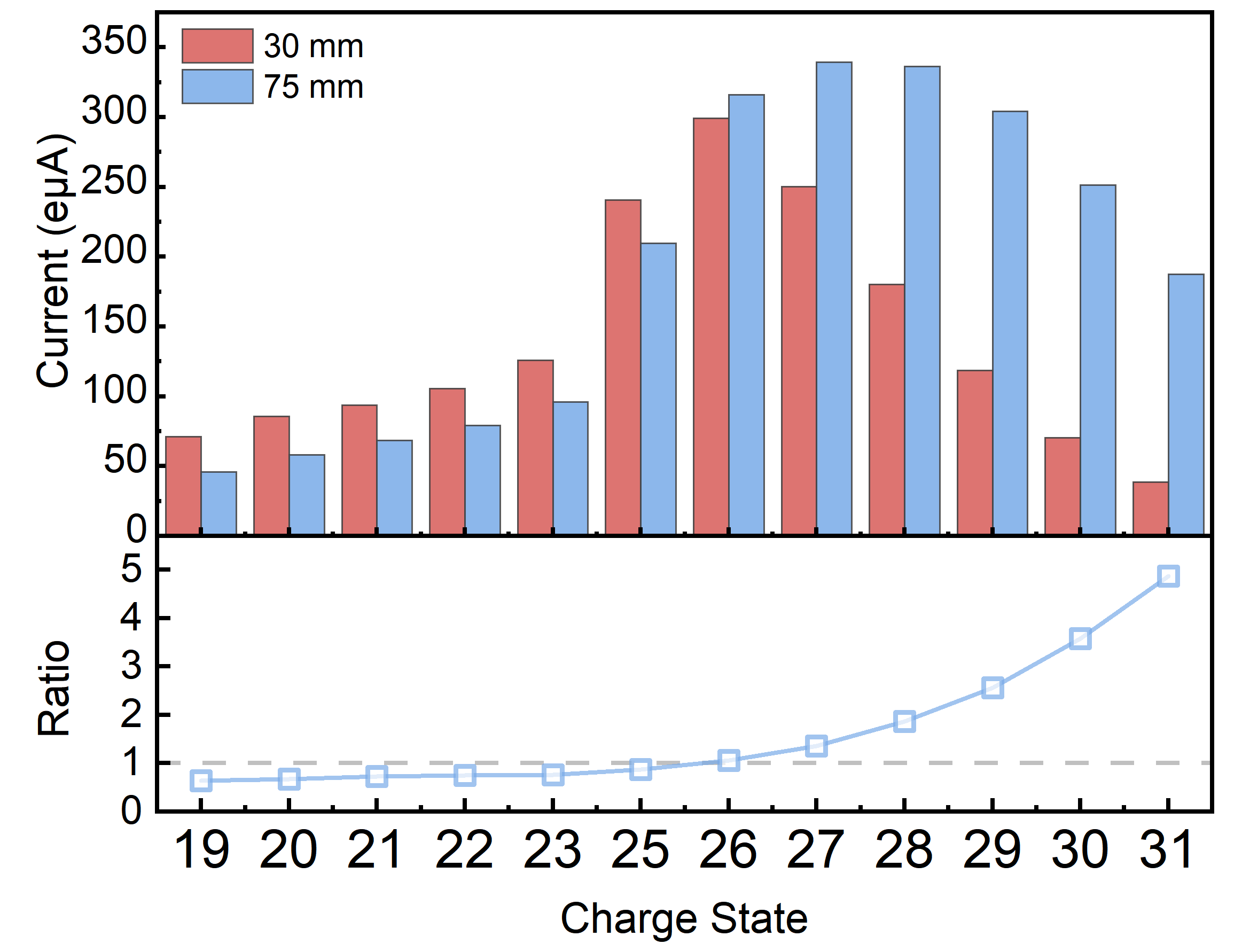}
    \caption{Xenon charge state distributions at the best and worst Vlasov launcher positions. Currents of  $\text{Xe}^{24+}$ are not shown because they overlap with $\text{O}^{3+}$. In addition, the gain factor of beam intensity at the best position compared to the worst position is presented.}
    \label{fig:Experimetal_Results_2}
\end{figure}

\section{Experimental results.}

To investigate the effect of the Vlasov launcher position on the extracted beam currents, we first tune the ion source at the out-most position ($z=0~\text{mm}$) to reach stable currents of the xenon ions with the specific charge state selected to measure. Then, the launcher position scan is done keeping all parameters constant. After the launcher is returned into initial position, the source is tuned for production of ions with another charge state and the procedure is repeated.~\autoref{fig:Experimetal_Results_1} presents the normalized beam intensities as a function of the Vlasov launcher position with different charge states, the microwave powers of $24~\text{GHz}$ and $18~\text{GHz}$ are fixed at $4.5$ and $0.5~\text{kW}$, respectively. It can be seen that the beam intensities initially decrease with the Vlasov launcher position until about $30~\text{mm}$, and then increase to their maximum values at $75~\text{mm}$. A further increase of the position leads to a small drop in the beam currents. Notably, the $45~\text{mm}$ distance between the optimum ($z=75~\text{mm}$) and worst ($z=30~\text{mm}$) positions far exceeds the $24~\text{GHz}$ microwave wavelength ($\lambda_{RF}=12.5~\text{mm}$),  suggesting that the observed behaviour is not due to standing wave formation. Meanwhile the beam currents obtained at the optimum position can be 4 times higher than in the worst case. Furthermore, the similar variation is also observed at different microwave power level (refer to supplementary material Figure~S2).

The effect of the Vlasov launcher position on the xenon charge state distribution (CSD) is also studied. The recorded m/q-spectra at the worst and best positions during the measurements of $\text{Xe}^{30+}$ currents shown in~\autoref{fig:Experimetal_Results_1} are chosen for display in~\autoref{fig:Experimetal_Results_2}. When transitioning from the worst to the optimum position, the CSD peak shifts towards higher charge state ($\text{Xe}^{26+}$ to $\text{Xe}^{27+}$). The beam intensity gain factor exceeds 1 for ions above the peak current CSD of the worst position ($\text{Xe}^{26+}$), and increases with the charge state, which indicates that optimization of the Vlasov launcher position is especially beneficial for highly charged ion production, similar to multiple frequency heating~\cite{10.1063/5.0076265}. Additionally, the total extracted ion current from the source reacts to the changes in the launcher position (see supplementary material Figure~S4) in parallel to the reaction of the highly charged ions.

\begin{figure}[!ht]
    \centering
    \includegraphics[width=\linewidth]{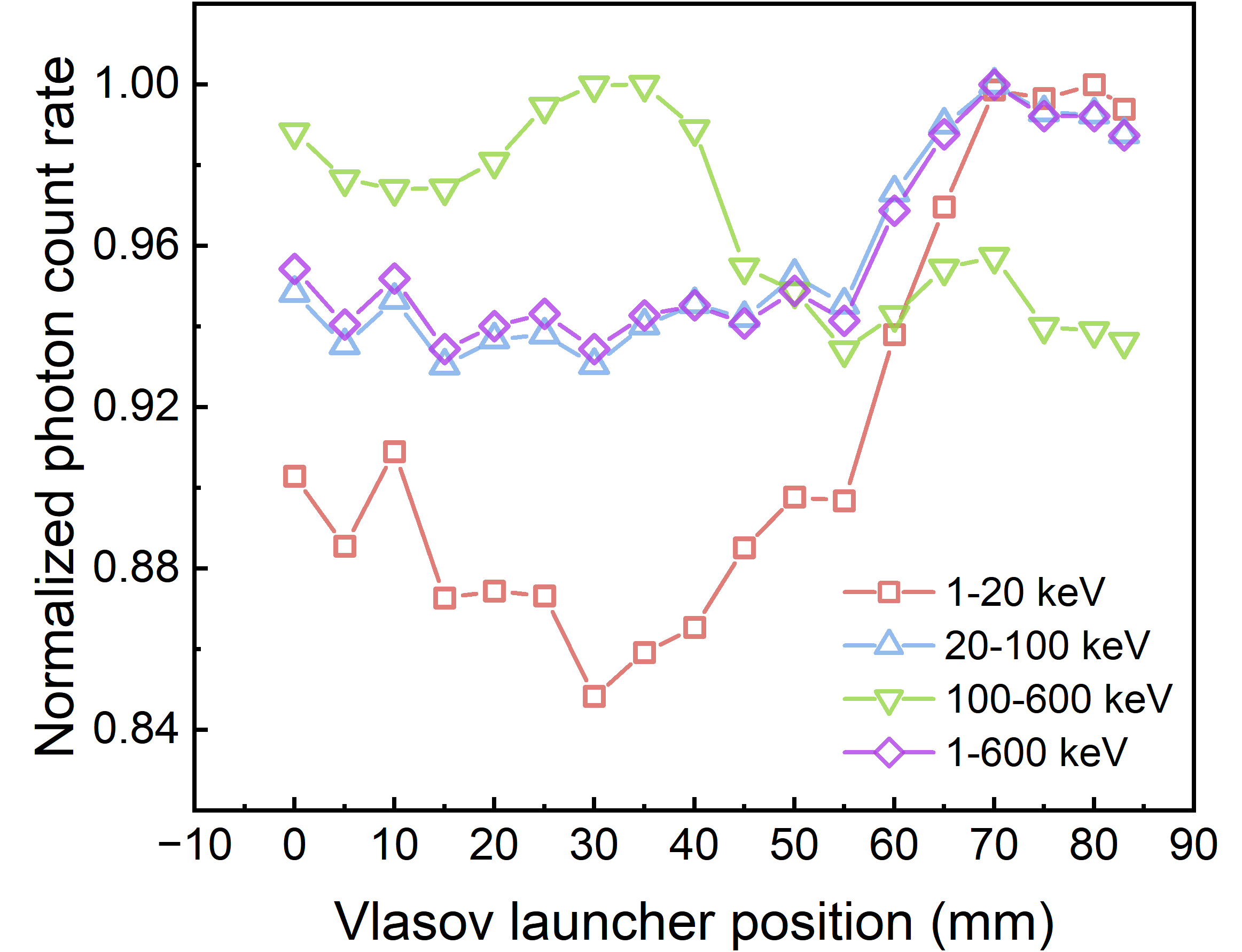}
    \caption{The normalized bremsstrahlung photon count rate within different energy intervals as a function of the Vlasov launcher position. The normalization is done to their corresponding maxima.}
    \label{fig:Experimetal_Results_3}
\end{figure}

The influence of the launcher position on the plasma electrons is studied indirectly by measuring the axial bremsstrahlung spectra.~\autoref{fig:Experimetal_Results_3} shows the normalized photon count rate within different energy intervals as a function of the launcher position. The corresponding bremsstrahlung spectra are synchronously obtained during the measurements of $\text{Xe}^{30+}$ beam currents shown in~\autoref{fig:Experimetal_Results_1} and corrected by the detector efficiency. It is found that the total count rate is increasing by $\sim$7\% in the optimized launcher position, indicating that higher plasma density are achieved. The $1\sim20~\text{keV}$ bremsstrahlung count rate is most affected by the launcher position, showing a tendency consistent with the beam current. The variation tendency of $20\sim100~\text{keV}$ interval is similar to that of $1\sim20~\text{keV}$, but the relative change is much smaller. On the other hand, the $100\sim600~\text{keV}$ count rate shows an almost opposite trend with that of $1\sim20~\text{keV}$, and reaches its maximum value at the worst position ($z=30~\text{mm}$) implying that more high energy electrons are produced. The original bremsstrahlung spectra at the best and worst positions can be found from the supplementary material Figure~S6, which supports the view of the launcher position affecting the plasma density.


\section{Numerical analysis.}

To better understand the experimental observations, based on the approach (details are shown in supplementary material, Section VII) developed in Refs.~\cite{Mironov_2020_Microwave,Microwaveabsorptionthesis}, we construct a full-size 3D model of the SECRAL-II ion source to calculate the microwave propagation and absorption with anisotropic dielectric tensor in the cold-plasma approximation by using the COMSOL Multiphysics\textsuperscript{\textregistered} RF module~\cite{COMSOL}. In our simulations, the microwave frequency and power are $24~\text{GHz}$ and $5~\text{kW}$, the 3D magnetic fields are calculated by the OPERA/Magnetostatic code using the coil current values used for tuning of $\text{Xe}^{30+}$ in~\autoref{fig:Experimetal_Results_1}, the electron density is set to $2.5\times 10^{12}~\text{cm}^{-3}$ where magnetic field magnitude $B \leq 1.1\times B_{ecr\_ {24~\text{GHz}}}$ (cold electron resonance field) and to 0 outside it. To avoid the solver's divergence, the total collision frequency of electrons with ions and neutral atoms is set to one thousandth of the microwave frequency, similar to Ref.~\cite{CLUGGISH201284}. 

 \begin{figure*}[!ht]
    \centering
    \includegraphics[width=0.9\linewidth]{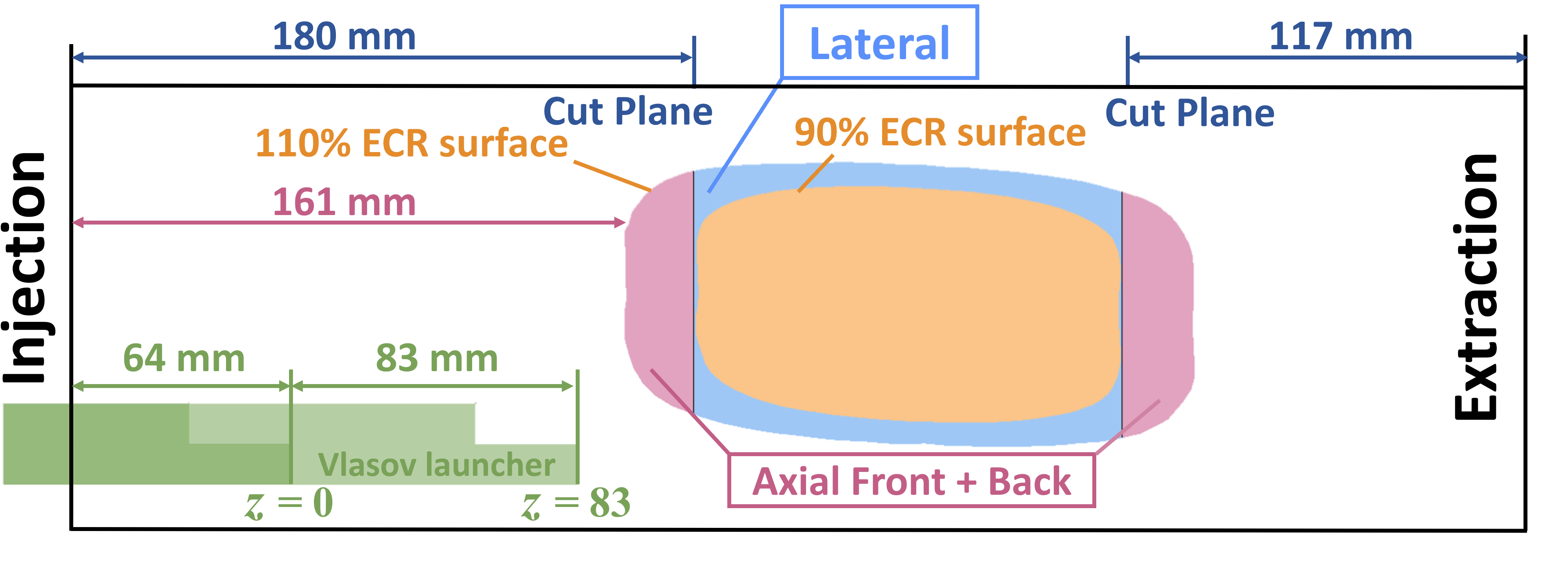}
    \caption{Computational domain and regions of interest in the COMSOL model. }
    \label{fig:Simulation_1}
\end{figure*}

\begin{figure}[!ht]
    \centering
    \includegraphics[width=\linewidth]{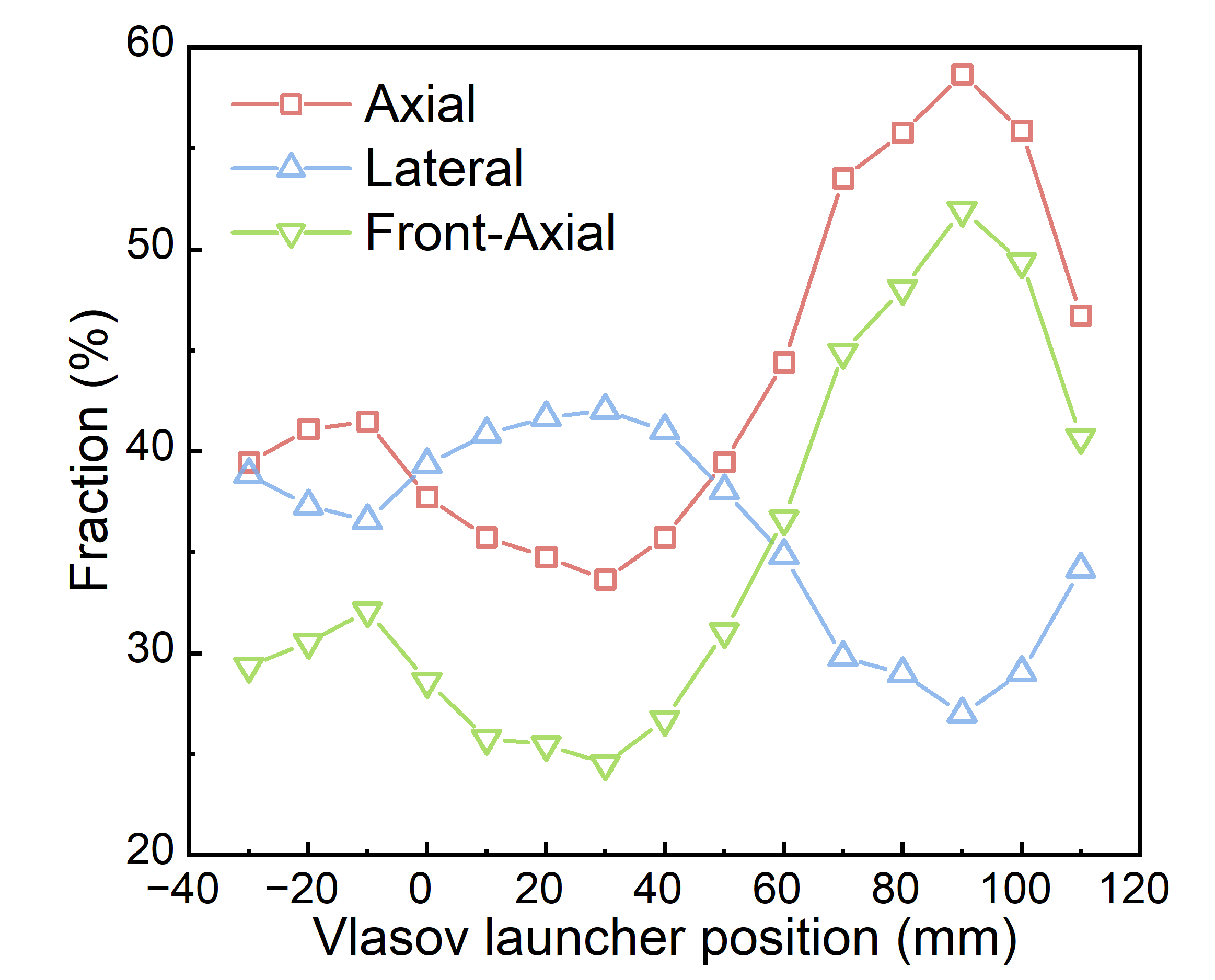}
    \caption{The fractional evolution in absorbed power with Vlasov Launcher position for axial, lateral and front-axial volumes.}
    \label{fig:Simulation_2}
\end{figure}

The experimental observations~\cite{LEITNER2005486} and numerical simulations~\cite{Mironov_2020} indicate that the ECR plasma is mostly localized close to the source axis inside the ECR zone. Since the wave propagation and absorption are strongly influenced by plasma density close to the ECR surface~\cite{CLUGGISH201284,3D-full_wave}, such localization affects the microwave coupling efficiency. To spatially characterize the microwave absorption process, we divide our computational domain into several regions of interest: the region close to the cold electron resonance surface of 24 GHz frequency (in the plasma volume where $0.9\times~B_{ecr\_24~\text{GHz}}~\leq~B~\leq~1.1\times~B_{ecr\_24~\text{GHz}}$), and the inner/outer parts for the rest of the domain. The ECR volume is additionally divided by two cutting planes into the axial-front, axial-back and the lateral parts as illustrated in~\autoref{fig:Simulation_1}. Absorption power evolution with Vlasov launcher position is calculated for axial (front + back), lateral, and front-axial volumes as shown in~\autoref{fig:Simulation_2}. According to the simulation model, the position of the Vlasov launcher affects significantly the absorbed power in the axial volume, especially in the front-axial volume. The relationship between absorbed power in the front-axial volume and launcher position aligns well with the experimental data of $\text{Xe}^{30+}$ beam intensity (see~\autoref{fig:Experimetal_Results_1}), with a maximum-to-minimum ratio of approximately 2. On the other hand, the absorbed power in the lateral volume shows an exactly opposite trend with that of front-axial volume, and reaches its maximum value at the worst position ($z=30~\text{mm}$), which coincides with the worst high charge state ion production.

\section{Discussion and Summary.} Experiments conducted with some $2^{nd}$ generation ECR ion sources~\cite{10.1063/1.2841694,10.1063/1.3665673,10.1063/1.4962026} have highlighted the influence of microwave frequency variations (frequency tuning) and cavity dimensions changes (cavity tuning) on extracted beam currents. It has been suggested~\cite{10.1063/1.2805665} that the beam intensity variations are caused by the excitation of electromagnetic eigenmodes within the plasma chamber acting as a resonator cavity. In $3^{rd}$ generation ECR ion sources, such as SECRAL-II with a larger plasma chamber diameter ($125~\text{mm}$) and higher frequency ($24~\text{GHz}$ with a $12.5~\text{mm}$ wavelength), the plasma chamber will act as an overmoded cavity in which the eigenmodes overlap due to plasma induced broadening. Then it can be expected~\cite{10.1063/1.4962026} that the frequency tuning and cavity tuning effects would be less effective. Indeed, we do not see in the reported measurements any changes in the source behavior that can be attributed to the cavity tuning with the expected oscillations when moving the launcher by around $\lambda_{RF}/2$. Strong gain in the source performance is observed when the Vlasov launcher is positioned such that the injected microwaves are focused on the ECR surface close the source axis -- currents of the highly charged xenon ions varies by a factor of about 4 and more ions with higher charge state are produced at the optimized position. This suggests that the microwave launching scheme plays an important role in the production of highly charged ion beams with superconducting ECR ion sources.

The recent numerical simulations~\cite{Mironov_2021,Mironov_2022} demonstrated that the ECR plasma can be characterized as a combination of the cold dense core and the hot dilute halo plasma, and the extracted ion currents are mostly defined by the core electron component, while the ion fluxes to the radial walls of the source are mostly produced in the halo plasma. From  the point of view of the microwave coupling efficiency, it is preferable to heat the core plasma  as much as possible and to reduce the power that is available for absorption by the halo electrons. According to our simulations, the axial volume corresponds to the region where the core plasma is supposed to be heated, we see that the variation of the absorbed power in the front-axial volume is strong and shows a good consistency with the extracted currents of high charge state ions, while the absorbed power in the lateral volume, which is believed to be responsible for the halo plasma heating, shows an opposite trend  with the beam currents. We argue that for a $3^{rd}$ generation ECR ion source, the location where the microwave power is absorbed is important for  highly charged ions production, and the optimum microwave absorption region is close to the source axis. The improved microwave heating efficiency could be achieved by the optimization of the microwave launching scheme, i.e., directing the microwaves to the ECR surface close to the source axis as much as possible, so that more plasma electrons located in the dense core regions can be effectively heated to the energies that are optimal for the highly charged ion production. This argument is supported by the diagnostic results: the evolution of the $100\sim600~\text{keV}$ bremsstrahlung count rate indicates that more high energy ($>100~\text{keV}$) electrons are produce at the worst position. As these high-energy  electrons also contribute to the $1\sim20~\text{keV}$ count rate, however, it is shown that the $1\sim20~\text{keV}$ count rate reaches a maximum at the best position. This should be due to more warm ($1\sim20~\text{keV}$) core electrons, which contribute part of the $1\sim20~\text{keV}$ count rate, and less high-energy electrons produced at the best position. As a result, the spectral temperature $T_{s}$, which can be considered as an indication of the average electron energy, reaches its minimum value at the best position, the evolution of $T_{s}$ with the Vlasov launcher position is presented in the supplementary material Figure~S7. Since these warm electrons are especially beneficial for the production of highly charged ions~\cite{7947187}, the performance of the ion source is thus improved.

\begin{table}[!ht]
    \caption{The optimized xenon beam intensity results using the movable Vlasov launcher}
    \centering
    \begin{ruledtabular}
    \begin{tabular}{cccc}
    Ion &  Beam intensity & Records& Improvement\\
    \colrule
    $\text{Xe}^{42+}$ &       $18~\text{e}\mu\text{A}$ &     $12~\text{e}\mu\text{A}$\footnote{$24+18~\text{GHz}$, $7\sim8~\text{kW}$, smaller circular waveguide.~\cite{PhysRevAccelBeams.25.063402,10.1063/1.5131101}} &     50   \% \\
    $\text{Xe}^{38+}$ &    $47~\text{e}\mu\text{A}$ &     $23~\text{e}\mu\text{A}^{\text{a}}$ &     104  \% \\
    $\text{Xe}^{34+}$ &    $146~\text{e}\mu\text{A}$ &     $120~\text{e}\mu\text{A}^{\text{a}}$ &     22   \% \\
    \end{tabular}
    \end{ruledtabular}
    \label{tab:The optimized xenon beam intensity results using the movable Vlasov launcher}
\end{table}

Based on these arguments, we completed a preliminary optimization of highly charged xenon ion beams at high microwave power. It is encouraging to see from~\autoref{tab:The optimized xenon beam intensity results using the movable Vlasov launcher} that intense highly charged xenon ion beams (e.g., $146~\text{e}\mu\text{A}$ of $\text{Xe}^{34+}$) could be obtained by using the movable Vlasov launcher. The improvement, compared with the beam intensity records of SECRAL ion source using smaller circular waveguide, is up to a factor of 2 (for $\text{Xe}^{38+}$) at the same power level. Meanwhile, the online test at high frequency and high microwave power has proved that the movable Vlasov launcher is reliable and could be a viable option for the routine operation.

In summary, the present work experimentally and numerically demonstrates that for a $3^{rd}$ generation ECR ion source operating at frequency above $20~\text{GHz}$, the microwave heating efficiency is significantly affected by the microwave launching scheme and high-efficiency microwave heating could be achieved by using the movable Vlasov launcher technique. The remarkable improvement of highly charged ion beam intensity based on this novel technique could directly and significantly enhance the performance of heavy ion accelerators and provide many new research opportunities in nuclear physics, atomic physics and other disciplines. Meanwhile, this study points out a direction for further study and optimization of microwave heating efficiency in modern superconducting ECR ion sources and thus gives a new insight into the design of microwave injection system for the next generation ECR ion sources.

\section{acknowledgments.}
The authors thank L. Celona and S. Gammino of INFN for helpful discussion, and D. Hitz for guidance on experimental design. This work has been supported by the National Natural Science Foundation of China (Grant Nos. 12025506, 11427904, and 12175285), the Key Scientific Instruments Development Program of CAS (Grant No. GJJSTD20210007),  and the Natural Science Foundation of Gansu Province, China (Grant No. 23JRRA584).



\providecommand{\noopsort}[1]{}\providecommand{\singleletter}[1]{#1}%


\begin{thebibliography}{36}%
\makeatletter
\providecommand \@ifxundefined [1]{%
 \@ifx{#1\undefined}
}%
\providecommand \@ifnum [1]{%
 \ifnum #1\expandafter \@firstoftwo
 \else \expandafter \@secondoftwo
 \fi
}%
\providecommand \@ifx [1]{%
 \ifx #1\expandafter \@firstoftwo
 \else \expandafter \@secondoftwo
 \fi
}%
\providecommand \natexlab [1]{#1}%
\providecommand \enquote  [1]{``#1''}%
\providecommand \bibnamefont  [1]{#1}%
\providecommand \bibfnamefont [1]{#1}%
\providecommand \citenamefont [1]{#1}%
\providecommand \href@noop [0]{\@secondoftwo}%
\providecommand \href [0]{\begingroup \@sanitize@url \@href}%
\providecommand \@href[1]{\@@startlink{#1}\@@href}%
\providecommand \@@href[1]{\endgroup#1\@@endlink}%
\providecommand \@sanitize@url [0]{\catcode `\\12\catcode `\$12\catcode
  `\&12\catcode `\#12\catcode `\^12\catcode `\_12\catcode `\%12\relax}%
\providecommand \@@startlink[1]{}%
\providecommand \@@endlink[0]{}%
\providecommand \url  [0]{\begingroup\@sanitize@url \@url }%
\providecommand \@url [1]{\endgroup\@href {#1}{\urlprefix }}%
\providecommand \urlprefix  [0]{URL }%
\providecommand \Eprint [0]{\href }%
\providecommand \doibase [0]{https://doi.org/}%
\providecommand \selectlanguage [0]{\@gobble}%
\providecommand \bibinfo  [0]{\@secondoftwo}%
\providecommand \bibfield  [0]{\@secondoftwo}%
\providecommand \translation [1]{[#1]}%
\providecommand \BibitemOpen [0]{}%
\providecommand \bibitemStop [0]{}%
\providecommand \bibitemNoStop [0]{.\EOS\space}%
\providecommand \EOS [0]{\spacefactor3000\relax}%
\providecommand \BibitemShut  [1]{\csname bibitem#1\endcsname}%
\let\auto@bib@innerbib\@empty
\bibitem [{\citenamefont {Geller}(1996)}]{Geller_book}%
  \BibitemOpen
  \bibfield  {author} {\bibinfo {author} {\bibfnamefont {R.}~\bibnamefont
  {Geller}},\ }\href {https://doi.org/https://doi.org/10.1201/9780203758663}
  {\emph {\bibinfo {title} {{{Electron Cyclotron Resonance Ion Sources and ECR
  Plasmas}}}}}\ (\bibinfo  {publisher} {Routledge},\ \bibinfo {year}
  {1996})\BibitemShut {NoStop}%
\bibitem [{\citenamefont {Zhou}\ \emph {et~al.}(2022)\citenamefont {Zhou},
  \citenamefont {Yang} \emph {et~al.}}]{Zhou2022}%
  \BibitemOpen
  \bibfield  {author} {\bibinfo {author} {\bibfnamefont {X.}~\bibnamefont
  {Zhou}}, \bibinfo {author} {\bibfnamefont {J.}~\bibnamefont {Yang}}, \emph
  {et~al.},\ }\href {https://doi.org/10.1007/s43673-022-00064-1} {\bibfield
  {journal} {\bibinfo  {journal} {AAPPS Bulletin}\ }\textbf {\bibinfo {volume}
  {32}},\ \bibinfo {pages} {35} (\bibinfo {year} {2022})}\BibitemShut {NoStop}%
\bibitem [{\citenamefont {Wei}\ \emph {et~al.}(2019)\citenamefont {Wei},
  \citenamefont {Ao}, \citenamefont {Beher} \emph
  {et~al.}}]{doi:10.1142/S0218301319300030}%
  \BibitemOpen
  \bibfield  {author} {\bibinfo {author} {\bibfnamefont {J.}~\bibnamefont
  {Wei}}, \bibinfo {author} {\bibfnamefont {H.}~\bibnamefont {Ao}}, \bibinfo
  {author} {\bibfnamefont {S.}~\bibnamefont {Beher}}, \emph {et~al.},\ }\href
  {https://doi.org/10.1142/S0218301319300030} {\bibfield  {journal} {\bibinfo
  {journal} {Int J Mod Phys E}\ }\textbf {\bibinfo {volume} {28}},\ \bibinfo
  {pages} {1930003} (\bibinfo {year} {2019})}\BibitemShut {NoStop}%
\bibitem [{\citenamefont {Delahaye}(2022)}]{franbergdelahaye:hiat2022-we2i1}%
  \BibitemOpen
  \bibfield  {author} {\bibinfo {author} {\bibfnamefont {H.~F.}\ \bibnamefont
  {Delahaye}},\ }in\ \href {https://doi.org/10.18429/JACoW-HIAT2022-WE2I1}
  {\emph {\bibinfo {booktitle} {Proc. HIAT'22}}}\ (\bibinfo {year} {2022})\
  pp.\ \bibinfo {pages} {124--129}\BibitemShut {NoStop}%
\bibitem [{\citenamefont {Okuno}\ \emph {et~al.}(2020)\citenamefont {Okuno},
  \citenamefont {Dantsuka}, \citenamefont {Fujimaki} \emph
  {et~al.}}]{Okuno_2020}%
  \BibitemOpen
  \bibfield  {author} {\bibinfo {author} {\bibfnamefont {H.}~\bibnamefont
  {Okuno}}, \bibinfo {author} {\bibfnamefont {T.}~\bibnamefont {Dantsuka}},
  \bibinfo {author} {\bibfnamefont {M.}~\bibnamefont {Fujimaki}}, \emph
  {et~al.},\ }\href {https://doi.org/10.1088/1742-6596/1401/1/012005}
  {\bibfield  {journal} {\bibinfo  {journal} {J. Phys.: Conf. Ser.}\ }\textbf
  {\bibinfo {volume} {1401}},\ \bibinfo {pages} {012005} (\bibinfo {year}
  {2020})}\BibitemShut {NoStop}%
\bibitem [{\citenamefont {Ostroumov}\ \emph {et~al.}(2021)\citenamefont
  {Ostroumov}, \citenamefont {Fukushima}, \citenamefont {Maruta}, \citenamefont
  {Plastun}, \citenamefont {Wei}, \citenamefont {Zhang},\ and\ \citenamefont
  {Zhao}}]{PhysRevLett.126.114801}%
  \BibitemOpen
  \bibfield  {author} {\bibinfo {author} {\bibfnamefont {P.~N.}\ \bibnamefont
  {Ostroumov}}, \bibinfo {author} {\bibfnamefont {K.}~\bibnamefont
  {Fukushima}}, \bibinfo {author} {\bibfnamefont {T.}~\bibnamefont {Maruta}},
  \bibinfo {author} {\bibfnamefont {A.~S.}\ \bibnamefont {Plastun}}, \bibinfo
  {author} {\bibfnamefont {J.}~\bibnamefont {Wei}}, \bibinfo {author}
  {\bibfnamefont {T.}~\bibnamefont {Zhang}},\ and\ \bibinfo {author}
  {\bibfnamefont {Q.}~\bibnamefont {Zhao}},\ }\href
  {https://doi.org/10.1103/PhysRevLett.126.114801} {\bibfield  {journal}
  {\bibinfo  {journal} {Phys. Rev. Lett.}\ }\textbf {\bibinfo {volume} {126}},\
  \bibinfo {pages} {114801} (\bibinfo {year} {2021})}\BibitemShut {NoStop}%
\bibitem [{\citenamefont {Khuyagbaatar}\ \emph {et~al.}(2020)\citenamefont
  {Khuyagbaatar}, \citenamefont {Yakushev}, \citenamefont {D\"ullmann} \emph
  {et~al.}}]{PhysRevC.102.064602}%
  \BibitemOpen
  \bibfield  {author} {\bibinfo {author} {\bibfnamefont {J.}~\bibnamefont
  {Khuyagbaatar}}, \bibinfo {author} {\bibfnamefont {A.}~\bibnamefont
  {Yakushev}}, \bibinfo {author} {\bibfnamefont {C.~E.}\ \bibnamefont
  {D\"ullmann}}, \emph {et~al.},\ }\href
  {https://doi.org/10.1103/PhysRevC.102.064602} {\bibfield  {journal} {\bibinfo
   {journal} {Phys. Rev. C}\ }\textbf {\bibinfo {volume} {102}},\ \bibinfo
  {pages} {064602} (\bibinfo {year} {2020})}\BibitemShut {NoStop}%
\bibitem [{\citenamefont {Celona}\ \emph {et~al.}(2010)\citenamefont {Celona},
  \citenamefont {Gammino}, \citenamefont {Ciavola} \emph
  {et~al.}}]{10.1063/1.3265366}%
  \BibitemOpen
  \bibfield  {author} {\bibinfo {author} {\bibfnamefont {L.}~\bibnamefont
  {Celona}}, \bibinfo {author} {\bibfnamefont {S.}~\bibnamefont {Gammino}},
  \bibinfo {author} {\bibfnamefont {G.}~\bibnamefont {Ciavola}}, \emph
  {et~al.},\ }\href {https://doi.org/10.1063/1.3265366} {\bibfield  {journal}
  {\bibinfo  {journal} {Rev. Sci. Instrum}\ }\textbf {\bibinfo {volume} {81}},\
  \bibinfo {pages} {02A333} (\bibinfo {year} {2010})}\BibitemShut {NoStop}%
\bibitem [{\citenamefont {Hitz}(2006)}]{Hitz2006RecentPI}%
  \BibitemOpen
  \bibfield  {author} {\bibinfo {author} {\bibfnamefont {D.}~\bibnamefont
  {Hitz}},\ }\href {https://api.semanticscholar.org/CorpusID:123397612}
  {\bibfield  {journal} {\bibinfo  {journal} {Adv Imag Elect Phys}\ }\textbf
  {\bibinfo {volume} {144}},\ \bibinfo {pages} {1} (\bibinfo {year}
  {2006})}\BibitemShut {NoStop}%
\bibitem [{\citenamefont {Gammino}\ \emph {et~al.}(2001)\citenamefont
  {Gammino}, \citenamefont {Ciavola}, \citenamefont {Celona} \emph
  {et~al.}}]{Gammino2002OperationsOT}%
  \BibitemOpen
  \bibfield  {author} {\bibinfo {author} {\bibfnamefont {S.}~\bibnamefont
  {Gammino}}, \bibinfo {author} {\bibfnamefont {G.}~\bibnamefont {Ciavola}},
  \bibinfo {author} {\bibfnamefont {L.}~\bibnamefont {Celona}}, \emph
  {et~al.},\ }\href {https://doi.org/10.1063/1.1435238} {\bibfield  {journal}
  {\bibinfo  {journal} {AIP Conf Proc}\ }\textbf {\bibinfo {volume} {600}},\
  \bibinfo {pages} {223} (\bibinfo {year} {2001})}\BibitemShut {NoStop}%
\bibitem [{\citenamefont {Marks}\ \emph {et~al.}(2005)\citenamefont {Marks},
  \citenamefont {Evans}, \citenamefont {Jory} \emph
  {et~al.}}]{10.1063/1.1893402}%
  \BibitemOpen
  \bibfield  {author} {\bibinfo {author} {\bibfnamefont {M.}~\bibnamefont
  {Marks}}, \bibinfo {author} {\bibfnamefont {S.}~\bibnamefont {Evans}},
  \bibinfo {author} {\bibfnamefont {H.}~\bibnamefont {Jory}}, \emph {et~al.},\
  }\href {https://doi.org/10.1063/1.1893402} {\bibfield  {journal} {\bibinfo
  {journal} {AIP Conf Proc}\ }\textbf {\bibinfo {volume} {749}},\ \bibinfo
  {pages} {207} (\bibinfo {year} {2005})}\BibitemShut {NoStop}%
\bibitem [{\citenamefont {Sun}\ \emph {et~al.}(2013)\citenamefont {Sun},
  \citenamefont {Lu}, \citenamefont {Feng} \emph {et~al.}}]{10.1063/1.4825164}%
  \BibitemOpen
  \bibfield  {author} {\bibinfo {author} {\bibfnamefont {L.}~\bibnamefont
  {Sun}}, \bibinfo {author} {\bibfnamefont {W.}~\bibnamefont {Lu}}, \bibinfo
  {author} {\bibfnamefont {Y.}~\bibnamefont {Feng}}, \emph {et~al.},\ }\href
  {https://doi.org/10.1063/1.4825164} {\bibfield  {journal} {\bibinfo
  {journal} {Rev. Sci. Instrum}\ }\textbf {\bibinfo {volume} {85}},\ \bibinfo
  {pages} {02A942} (\bibinfo {year} {2013})}\BibitemShut {NoStop}%
\bibitem [{\citenamefont {Nakagawa}\ \emph {et~al.}(2010)\citenamefont
  {Nakagawa}, \citenamefont {Higurashi}, \citenamefont {Ohnishi} \emph
  {et~al.}}]{Riken}%
  \BibitemOpen
  \bibfield  {author} {\bibinfo {author} {\bibfnamefont {T.}~\bibnamefont
  {Nakagawa}}, \bibinfo {author} {\bibfnamefont {Y.}~\bibnamefont {Higurashi}},
  \bibinfo {author} {\bibfnamefont {J.}~\bibnamefont {Ohnishi}}, \emph
  {et~al.},\ }\href {https://doi.org/10.1063/1.3259232} {\bibfield  {journal}
  {\bibinfo  {journal} {Rev. Sci. Instrum}\ }\textbf {\bibinfo {volume} {81}},\
  \bibinfo {pages} {02A320} (\bibinfo {year} {2010})}\BibitemShut {NoStop}%
\bibitem [{\citenamefont {Zhao}\ \emph {et~al.}(2018)\citenamefont {Zhao},
  \citenamefont {Sun}, \citenamefont {Guo} \emph {et~al.}}]{10.1063/1.5017479}%
  \BibitemOpen
  \bibfield  {author} {\bibinfo {author} {\bibfnamefont {H.}~\bibnamefont
  {Zhao}}, \bibinfo {author} {\bibfnamefont {L.}~\bibnamefont {Sun}}, \bibinfo
  {author} {\bibfnamefont {J.}~\bibnamefont {Guo}}, \emph {et~al.},\ }\href
  {https://doi.org/10.1063/1.5017479} {\bibfield  {journal} {\bibinfo
  {journal} {Rev. Sci. Instrum}\ }\textbf {\bibinfo {volume} {89}},\ \bibinfo
  {pages} {052301} (\bibinfo {year} {2018})}\BibitemShut {NoStop}%
\bibitem [{\citenamefont {Lyneis}\ \emph {et~al.}(2013)\citenamefont {Lyneis},
  \citenamefont {Benitez}, \citenamefont {Hodgkinson} \emph
  {et~al.}}]{10.1063/1.4832064}%
  \BibitemOpen
  \bibfield  {author} {\bibinfo {author} {\bibfnamefont {C.}~\bibnamefont
  {Lyneis}}, \bibinfo {author} {\bibfnamefont {J.}~\bibnamefont {Benitez}},
  \bibinfo {author} {\bibfnamefont {A.}~\bibnamefont {Hodgkinson}}, \emph
  {et~al.},\ }\href {https://doi.org/10.1063/1.4832064} {\bibfield  {journal}
  {\bibinfo  {journal} {Rev. Sci. Instrum}\ }\textbf {\bibinfo {volume} {85}},\
  \bibinfo {pages} {02A932} (\bibinfo {year} {2013})}\BibitemShut {NoStop}%
\bibitem [{\citenamefont {Sun}\ \emph {et~al.}(2015{\natexlab{a}})\citenamefont
  {Sun}, \citenamefont {Guo}, \citenamefont {Lu} \emph
  {et~al.}}]{Sun2016AdvancementOH}%
  \BibitemOpen
  \bibfield  {author} {\bibinfo {author} {\bibfnamefont {L.}~\bibnamefont
  {Sun}}, \bibinfo {author} {\bibfnamefont {J.}~\bibnamefont {Guo}}, \bibinfo
  {author} {\bibfnamefont {W.}~\bibnamefont {Lu}}, \emph {et~al.},\ }\href
  {https://doi.org/10.1063/1.4933123} {\bibfield  {journal} {\bibinfo
  {journal} {Rev. Sci. Instrum}\ }\textbf {\bibinfo {volume} {87}},\ \bibinfo
  {pages} {02A707} (\bibinfo {year} {2015}{\natexlab{a}})}\BibitemShut
  {NoStop}%
\bibitem [{\citenamefont {Guo}\ \emph {et~al.}(2020)\citenamefont {Guo},
  \citenamefont {Sun}, \citenamefont {Lu} \emph {et~al.}}]{10.1063/1.5131101}%
  \BibitemOpen
  \bibfield  {author} {\bibinfo {author} {\bibfnamefont {J.}~\bibnamefont
  {Guo}}, \bibinfo {author} {\bibfnamefont {L.}~\bibnamefont {Sun}}, \bibinfo
  {author} {\bibfnamefont {W.}~\bibnamefont {Lu}}, \emph {et~al.},\ }\href
  {https://doi.org/10.1063/1.5131101} {\bibfield  {journal} {\bibinfo
  {journal} {Rev. Sci. Instrum}\ }\textbf {\bibinfo {volume} {91}},\ \bibinfo
  {pages} {013322} (\bibinfo {year} {2020})}\BibitemShut {NoStop}%
\bibitem [{\citenamefont {Vlasov}\ and\ \citenamefont
  {Orlova}(1974)}]{10.1007/BF01037072}%
  \BibitemOpen
  \bibfield  {author} {\bibinfo {author} {\bibfnamefont {S.}~\bibnamefont
  {Vlasov}}\ and\ \bibinfo {author} {\bibfnamefont {I.}~\bibnamefont
  {Orlova}},\ }\href {https://doi.org/10.1007/BF01037072} {\bibfield  {journal}
  {\bibinfo  {journal} {Radiophys. Quantum Electron.}\ }\textbf {\bibinfo
  {volume} {17}},\ \bibinfo {pages} {115–119} (\bibinfo {year}
  {1974})}\BibitemShut {NoStop}%
\bibitem [{\citenamefont {Li}\ \emph {et~al.}(2022)\citenamefont {Li},
  \citenamefont {Li}, \citenamefont {Ma} \emph
  {et~al.}}]{PhysRevAccelBeams.25.063402}%
  \BibitemOpen
  \bibfield  {author} {\bibinfo {author} {\bibfnamefont {L.~X.}\ \bibnamefont
  {Li}}, \bibinfo {author} {\bibfnamefont {J.~B.}\ \bibnamefont {Li}}, \bibinfo
  {author} {\bibfnamefont {J.~D.}\ \bibnamefont {Ma}}, \emph {et~al.},\ }\href
  {https://doi.org/10.1103/PhysRevAccelBeams.25.063402} {\bibfield  {journal}
  {\bibinfo  {journal} {Phys. Rev. Accel. Beams}\ }\textbf {\bibinfo {volume}
  {25}},\ \bibinfo {pages} {063402} (\bibinfo {year} {2022})}\BibitemShut
  {NoStop}%
\bibitem [{\citenamefont {Sun}\ \emph {et~al.}(2015{\natexlab{b}})\citenamefont
  {Sun}, \citenamefont {Lu}, \citenamefont {Wu} \emph
  {et~al.}}]{Sun2015STATUSRO}%
  \BibitemOpen
  \bibfield  {author} {\bibinfo {author} {\bibfnamefont {L.}~\bibnamefont
  {Sun}}, \bibinfo {author} {\bibfnamefont {W.}~\bibnamefont {Lu}}, \bibinfo
  {author} {\bibfnamefont {W.}~\bibnamefont {Wu}}, \emph {et~al.},\ }in\ \href
  {https://api.semanticscholar.org/CorpusID:202623770} {\emph {\bibinfo
  {booktitle} {Proceedings of ECRIS-2014}}}\ (\bibinfo {year} {2015})\ pp.\
  \bibinfo {pages} {94--98}\BibitemShut {NoStop}%
\bibitem [{\citenamefont {Li}\ \emph {et~al.}(2020)\citenamefont {Li},
  \citenamefont {Li}, \citenamefont {Bhaskar} \emph
  {et~al.}}]{WOS:000560067700001}%
  \BibitemOpen
  \bibfield  {author} {\bibinfo {author} {\bibfnamefont {J.}~\bibnamefont
  {Li}}, \bibinfo {author} {\bibfnamefont {L.}~\bibnamefont {Li}}, \bibinfo
  {author} {\bibfnamefont {B.~S.}\ \bibnamefont {Bhaskar}}, \emph {et~al.},\
  }\href {https://doi.org/10.1088/1361-6587/ab9d8f} {\bibfield  {journal}
  {\bibinfo  {journal} {Plasma Phys. Control. Fusion}\ }\textbf {\bibinfo
  {volume} {62}},\ \bibinfo {pages} {095015} (\bibinfo {year}
  {2020})}\BibitemShut {NoStop}%
\bibitem [{\citenamefont {Vondrasek}(2022)}]{10.1063/5.0076265}%
  \BibitemOpen
  \bibfield  {author} {\bibinfo {author} {\bibfnamefont {R.}~\bibnamefont
  {Vondrasek}},\ }\href {https://doi.org/10.1063/5.0076265} {\bibfield
  {journal} {\bibinfo  {journal} {Rev. Sci. Instrum}\ }\textbf {\bibinfo
  {volume} {93}},\ \bibinfo {pages} {031501} (\bibinfo {year}
  {2022})}\BibitemShut {NoStop}%
\bibitem [{\citenamefont {Mironov}\ \emph
  {et~al.}(2020{\natexlab{a}})\citenamefont {Mironov}, \citenamefont
  {Bogomolov}, \citenamefont {Bondarchenko} \emph
  {et~al.}}]{Mironov_2020_Microwave}%
  \BibitemOpen
  \bibfield  {author} {\bibinfo {author} {\bibfnamefont {V.}~\bibnamefont
  {Mironov}}, \bibinfo {author} {\bibfnamefont {S.}~\bibnamefont {Bogomolov}},
  \bibinfo {author} {\bibfnamefont {A.}~\bibnamefont {Bondarchenko}}, \emph
  {et~al.},\ }\href {https://doi.org/10.1088/1748-0221/15/10/P10030} {\bibfield
   {journal} {\bibinfo  {journal} {Journal of Instrumentation}\ }\textbf
  {\bibinfo {volume} {15}}\bibinfo  {number} { (10)},\ \bibinfo {pages}
  {P10030}}\BibitemShut {NoStop}%
\bibitem [{\citenamefont {van Ameijde}(2021)}]{Microwaveabsorptionthesis}%
  \BibitemOpen
\bibfield  {number} {  }\bibfield  {author} {\bibinfo {author} {\bibfnamefont
  {D.~A.}\ \bibnamefont {van Ameijde}},\ }\emph {\bibinfo {title} {{Microwave
  absorption in an ECR plasma: A theoretical analysis and computational
  implementation for unstructured meshes using vector finite elements}}},\
  \href
  {https://research.tue.nl/files/188052773/0822307_Ameijde_D.A._van_MSc_thesis_MAP_VERTROUWELIJK_TM_1_2_2022.pdf}
  {Master's thesis},\ \bibinfo  {school} {Eindhoven University of Technology}
  (\bibinfo {year} {2021})\BibitemShut {NoStop}%
\bibitem [{COMSOL Multiphysics\textsuperscript{\textregistered} Version
  6.1()}]{COMSOL}%
  \BibitemOpen
  COMSOL Multiphysics\textsuperscript{\textregistered} Version 6.1,\ \href
  {https://www.comsol.com/rf-module} {\emph {\bibinfo {title} {{COMSOL
  Multiphysics\textsuperscript{\textregistered} RF module}}}}\ (\bibinfo {year}
  {2022})\BibitemShut {NoStop}%
\bibitem [{\citenamefont {Cluggish}\ and\ \citenamefont
  {Kim}(2012)}]{CLUGGISH201284}%
  \BibitemOpen
  \bibfield  {author} {\bibinfo {author} {\bibfnamefont {B.~P.}\ \bibnamefont
  {Cluggish}}\ and\ \bibinfo {author} {\bibfnamefont {J.~S.}\ \bibnamefont
  {Kim}},\ }\href {https://doi.org/https://doi.org/10.1016/j.nima.2011.10.015}
  {\bibfield  {journal} {\bibinfo  {journal} {Nucl. Instrum. Methods. Phys.
  Res. A}\ }\textbf {\bibinfo {volume} {664}},\ \bibinfo {pages} {84} (\bibinfo
  {year} {2012})}\BibitemShut {NoStop}%
\bibitem [{\citenamefont {Leitner}\ \emph {et~al.}(2005)\citenamefont
  {Leitner}, \citenamefont {Lyneis}, \citenamefont {Abbott} \emph
  {et~al.}}]{LEITNER2005486}%
  \BibitemOpen
  \bibfield  {author} {\bibinfo {author} {\bibfnamefont {D.}~\bibnamefont
  {Leitner}}, \bibinfo {author} {\bibfnamefont {C.}~\bibnamefont {Lyneis}},
  \bibinfo {author} {\bibfnamefont {S.}~\bibnamefont {Abbott}}, \emph
  {et~al.},\ }\href
  {https://doi.org/https://doi.org/10.1016/j.nimb.2005.03.230} {\bibfield
  {journal} {\bibinfo  {journal} {Nucl Instrum Methods Phys Res B}\ }\textbf
  {\bibinfo {volume} {235}},\ \bibinfo {pages} {486} (\bibinfo {year}
  {2005})}\BibitemShut {NoStop}%
\bibitem [{\citenamefont {Mironov}\ \emph
  {et~al.}(2020{\natexlab{b}})\citenamefont {Mironov}, \citenamefont
  {Bogomolov}, \citenamefont {Bondarchenko} \emph {et~al.}}]{Mironov_2020}%
  \BibitemOpen
  \bibfield  {author} {\bibinfo {author} {\bibfnamefont {V.}~\bibnamefont
  {Mironov}}, \bibinfo {author} {\bibfnamefont {S.}~\bibnamefont {Bogomolov}},
  \bibinfo {author} {\bibfnamefont {A.}~\bibnamefont {Bondarchenko}}, \emph
  {et~al.},\ }\href {https://doi.org/10.1088/1361-6595/ab62dc} {\bibfield
  {journal} {\bibinfo  {journal} {Plasma Sources Sci Technol}\ }\textbf
  {\bibinfo {volume} {29}},\ \bibinfo {pages} {065010} (\bibinfo {year}
  {2020}{\natexlab{b}})}\BibitemShut {NoStop}%
\bibitem [{\citenamefont {Mascali}\ \emph {et~al.}(2015)\citenamefont
  {Mascali}, \citenamefont {Torrisi}, \citenamefont {Neri} \emph
  {et~al.}}]{3D-full_wave}%
  \BibitemOpen
  \bibfield  {author} {\bibinfo {author} {\bibfnamefont {D.}~\bibnamefont
  {Mascali}}, \bibinfo {author} {\bibfnamefont {G.}~\bibnamefont {Torrisi}},
  \bibinfo {author} {\bibfnamefont {L.}~\bibnamefont {Neri}}, \emph {et~al.},\
  }\href {https://doi.org/10.1140/epjd/e2014-50168-5} {\bibfield  {journal}
  {\bibinfo  {journal} {Eur. Phys. J. D}\ }\textbf {\bibinfo {volume} {69}},\
  \bibinfo {pages} {27} (\bibinfo {year} {2015})}\BibitemShut {NoStop}%
\bibitem [{\citenamefont {Celona}\ \emph {et~al.}(2008)\citenamefont {Celona},
  \citenamefont {Ciavola}, \citenamefont {Consoli} \emph
  {et~al.}}]{10.1063/1.2841694}%
  \BibitemOpen
  \bibfield  {author} {\bibinfo {author} {\bibfnamefont {L.}~\bibnamefont
  {Celona}}, \bibinfo {author} {\bibfnamefont {G.}~\bibnamefont {Ciavola}},
  \bibinfo {author} {\bibfnamefont {F.}~\bibnamefont {Consoli}}, \emph
  {et~al.},\ }\href {https://doi.org/10.1063/1.2841694} {\bibfield  {journal}
  {\bibinfo  {journal} {Rev. Sci. Instrum}\ }\textbf {\bibinfo {volume} {79}},\
  \bibinfo {pages} {023305} (\bibinfo {year} {2008})}\BibitemShut {NoStop}%
\bibitem [{\citenamefont {Maimone}\ \emph {et~al.}(2011)\citenamefont
  {Maimone}, \citenamefont {Celona}, \citenamefont {Lang} \emph
  {et~al.}}]{10.1063/1.3665673}%
  \BibitemOpen
  \bibfield  {author} {\bibinfo {author} {\bibfnamefont {F.}~\bibnamefont
  {Maimone}}, \bibinfo {author} {\bibfnamefont {L.}~\bibnamefont {Celona}},
  \bibinfo {author} {\bibfnamefont {R.}~\bibnamefont {Lang}}, \emph {et~al.},\
  }\href {https://doi.org/10.1063/1.3665673} {\bibfield  {journal} {\bibinfo
  {journal} {Rev. Sci. Instrum}\ }\textbf {\bibinfo {volume} {82}},\ \bibinfo
  {pages} {123302} (\bibinfo {year} {2011})}\BibitemShut {NoStop}%
\bibitem [{\citenamefont {Tarvainen}\ \emph {et~al.}(2016)\citenamefont
  {Tarvainen}, \citenamefont {Orpana}, \citenamefont {Kronholm} \emph
  {et~al.}}]{10.1063/1.4962026}%
  \BibitemOpen
  \bibfield  {author} {\bibinfo {author} {\bibfnamefont {O.}~\bibnamefont
  {Tarvainen}}, \bibinfo {author} {\bibfnamefont {J.}~\bibnamefont {Orpana}},
  \bibinfo {author} {\bibfnamefont {R.}~\bibnamefont {Kronholm}}, \emph
  {et~al.},\ }\href {https://doi.org/10.1063/1.4962026} {\bibfield  {journal}
  {\bibinfo  {journal} {Rev. Sci. Instrum}\ }\textbf {\bibinfo {volume} {87}},\
  \bibinfo {pages} {093301} (\bibinfo {year} {2016})}\BibitemShut {NoStop}%
\bibitem [{\citenamefont {Consoli}\ \emph {et~al.}(2008)\citenamefont
  {Consoli}, \citenamefont {Celona}, \citenamefont {Ciavola} \emph
  {et~al.}}]{10.1063/1.2805665}%
  \BibitemOpen
  \bibfield  {author} {\bibinfo {author} {\bibfnamefont {F.}~\bibnamefont
  {Consoli}}, \bibinfo {author} {\bibfnamefont {L.}~\bibnamefont {Celona}},
  \bibinfo {author} {\bibfnamefont {G.}~\bibnamefont {Ciavola}}, \emph
  {et~al.},\ }\href {https://doi.org/10.1063/1.2805665} {\bibfield  {journal}
  {\bibinfo  {journal} {Rev. Sci. Instrum}\ }\textbf {\bibinfo {volume} {79}},\
  \bibinfo {pages} {02A308} (\bibinfo {year} {2008})}\BibitemShut {NoStop}%
\bibitem [{\citenamefont {Mironov}\ \emph {et~al.}(2021)\citenamefont
  {Mironov}, \citenamefont {Bogomolov}, \citenamefont {Bondarchenko} \emph
  {et~al.}}]{Mironov_2021}%
  \BibitemOpen
  \bibfield  {author} {\bibinfo {author} {\bibfnamefont {V.}~\bibnamefont
  {Mironov}}, \bibinfo {author} {\bibfnamefont {S.}~\bibnamefont {Bogomolov}},
  \bibinfo {author} {\bibfnamefont {A.}~\bibnamefont {Bondarchenko}}, \emph
  {et~al.},\ }\href {https://doi.org/10.1088/1748-0221/16/04/P04009} {\bibfield
   {journal} {\bibinfo  {journal} {J. Instrum}\ }\textbf {\bibinfo {volume}
  {16}},\ \bibinfo {pages} {P04009} (\bibinfo {year} {2021})}\BibitemShut
  {NoStop}%
\bibitem [{\citenamefont {Mironov}\ \emph {et~al.}(2022)\citenamefont
  {Mironov}, \citenamefont {Bogomolov}, \citenamefont {Bondarchenko} \emph
  {et~al.}}]{Mironov_2022}%
  \BibitemOpen
  \bibfield  {author} {\bibinfo {author} {\bibfnamefont {V.}~\bibnamefont
  {Mironov}}, \bibinfo {author} {\bibfnamefont {S.}~\bibnamefont {Bogomolov}},
  \bibinfo {author} {\bibfnamefont {A.}~\bibnamefont {Bondarchenko}}, \emph
  {et~al.},\ }\href {https://doi.org/10.1088/1748-0221/17/06/P06028} {\bibfield
   {journal} {\bibinfo  {journal} {J. Instrum}\ }\textbf {\bibinfo {volume}
  {17}},\ \bibinfo {pages} {P06028} (\bibinfo {year} {2022})}\BibitemShut
  {NoStop}%
\bibitem [{\citenamefont {Benitez}\ \emph {et~al.}(2017)\citenamefont
  {Benitez}, \citenamefont {Lyneis}, \citenamefont {Phair} \emph
  {et~al.}}]{7947187}%
  \BibitemOpen
  \bibfield  {author} {\bibinfo {author} {\bibfnamefont {J.}~\bibnamefont
  {Benitez}}, \bibinfo {author} {\bibfnamefont {C.}~\bibnamefont {Lyneis}},
  \bibinfo {author} {\bibfnamefont {L.}~\bibnamefont {Phair}}, \emph {et~al.},\
  }\href {https://doi.org/10.1109/TPS.2017.2706718} {\bibfield  {journal}
  {\bibinfo  {journal} {IEEE Trans Plasma Sci}\ }\textbf {\bibinfo {volume}
  {45}},\ \bibinfo {pages} {1746} (\bibinfo {year} {2017})}\BibitemShut
  {NoStop}%
\end{thebibliography}
\end{document}